\documentclass[conference]{IEEEtran}
\IEEEoverridecommandlockouts
\usepackage{cite}
\usepackage{amsmath,amssymb,amsfonts}
\usepackage{algorithmic}
\usepackage{graphicx}
\usepackage{textcomp}
\usepackage{xcolor}
\usepackage{subcaption}
\def\BibTeX{{\rm B\kern-.05em{\sc i\kern-.025em b}\kern-.08em
    T\kern-.1667em\lower.7ex\hbox{E}\kern-.125emX}}
\begin{document}

\title{Understanding Cyber Athletes Behaviour \\Through a Smart Chair: \\CS:GO and Monolith Team Scenario
\thanks{The reported study was funded by RFBR according to the research project No. 18-29-22077\textbackslash18.}
}
%Anton Smerdov1, Anastasia Kiskun1,2, Rostislav Shaniiazov1,2, Andrey Somov1,*, Evgeny Burnaev1

\author{
	\IEEEauthorblockN{
		Anton Smerdov\IEEEauthorrefmark{1}\IEEEauthorrefmark{2},
		Anastasia Kiskun\IEEEauthorrefmark{1}\IEEEauthorrefmark{3},
		Rostislav Shaniiazov\IEEEauthorrefmark{1}\IEEEauthorrefmark{3},
		Andrey Somov\IEEEauthorrefmark{1}\IEEEauthorrefmark{4},
		Evgeny Burnaev\IEEEauthorrefmark{1}}
	\IEEEauthorblockA{
		\IEEEauthorrefmark{1}Skolkovo Institute of Science and Technology, CDISE, Moscow, Russia}
	\IEEEauthorblockA{
		\IEEEauthorrefmark{2}Moscow Institute of Physics and Technology, Moscow, Russia}
	\IEEEauthorblockA{
		\IEEEauthorrefmark{3}Saint Petersburg State University of Aerospace Instrumentation, Saint Petersburg, Russia}
	\IEEEauthorblockA{
		\IEEEauthorrefmark{4}a.somov@skoltech.ru}
	}

\IEEEoverridecommandlockouts
\IEEEpubid{\makebox[\columnwidth]{978-1-5386-4980-0/19/\$31.00 \copyright2019 IEEE} \hspace{\columnsep}\makebox[\columnwidth]{ }}

\maketitle

%\author{\IEEEauthorblockN{Anton Smerdov}
%\IEEEauthorblockA{\textit{Skolkovo Institute of Science and Technology} \\
%\textit{CDISE}\\
%Moscow, Russia \\
%Anton.Smerdov@skoltech.ru}
%\and
%\IEEEauthorblockN{Anastasia Kiskun}
%\IEEEauthorblockA{\textit{Skolkovo Institute of Science and Technology} \\
%\textit{CDISE}\\
%Moscow, Russia \\
%Anastasia.Kishkun@skolkovotech.ru}
%\and
%\IEEEauthorblockN{Rostislav Shaniiazov}
%\IEEEauthorblockA{\textit{Skolkovo Institute of Science and Technology} \\
%\textit{CDISE}\\
%Moscow, Russia \\
%Rostislav.Shaniiazov@skoltech.ru}
%\and
%\IEEEauthorblockN{Andrey Somov}
%\IEEEauthorblockA{\textit{Skolkovo Institute of Science and Technology} \\
%\textit{CDISE}\\
%Moscow, Russia \\
%a.somov@skoltech.ru}
%\and
%\IEEEauthorblockN{Evgeny Burnaev}
%\IEEEauthorblockA{\textit{Skolkovo Institute of Science and Technology} \\
%\textit{CDISE}\\
%Moscow, Russia \\
%e.burnaev@skoltech.ru}
%}

\begin{abstract}
eSports is the rapidly developing multidisciplinary domain. However, research and experimentation in eSports are in the infancy. In this work, we propose a smart chair platform - an unobtrusive approach to the collection of data on the eSports athletes and data further processing with machine learning methods. The use case scenario involves three groups of players: ‘cyber athletes’ (Monolith team), semi-professional players and newbies all playing CS:GO discipline. In particular, we collect data from the accelerometer and gyroscope integrated in the chair and apply machine learning algorithms for the data analysis. Our results demonstrate that the professional athletes can be identified by their behaviour on the chair while playing the game.
\end{abstract}

\begin{IEEEkeywords}
eSports, smart sensing, machine learning
\end{IEEEkeywords}

\section{Introduction}

Within the last few years cybersport (or eSports) has progressed from the video gaming entertainment to the sport domain professionally recognised in many countries. eSports has become a self-sustainable industry involving professional teams, their managers and coaches. In terms of training activities, there is still a lack of tools aimed at enhancing  professional skills and providing valuable feedback on the demonstrated performance. Indeed, a number of online services provide the player with a feedback based on game statistics. At the same time, most of the eSports team coaches rely on their experience rather than science while planning the training process. Collecting more data is to open wide vista for  optimization of the training process and devising game strategies based on the personal feedback.

A straightforward approach is to equip the players with extra wearable sensors or sensors integrated in a chair and perform further data analysis using powerful cloud facilities: on top of analyzing each particular player, there is an option for predicting his playing strategy in a particular situation, e.g. \textit{clutch} in \textit{Counter Strike:Global Offensive} (CS:GO) discipline.

At the moment research in eSports is in its infancy. To the best of our knowledge there are no recent works trying to combine the prediction of a cyber athlete performance by his behaviour relying on wearable sensors or sensors integrated in a chair. However, a smart chair concept has already been  proposed in a number of research papers. For example, authors
in \cite{smart_chair} report on the posture detection using the tilt sensors in addition to the pressure sensors. Kumar et al.  \cite{chair_activity_1} used data from the pressure sensors embedded in a smart chair for classification of functional activities, e.g. talking, coughing, eating, and emotion based activities, e.g., crying, laughing, shouting. The authors also managed to estimate breathing rate during the static sedentary postures. Pressure sensors are extensively used in \cite{smart_chair_pressure_1, smart_chair_pressure_2} for identifying the sedentary patterns. In \cite{chair_stress} the pressure sensors are used for predicting the stress level of a user and in \cite{chair_activity_2} for predicting the subsequent sitting activity.
A smart chair is also applied to measuring the vital health signs like the heart rate through the Ballistocardiograhy (BCG) technique, Junnila et al. used EMFi-film sensor installed on the seat of the chair \cite{chair_bcg}. Also, ECG can be computed by the sensors embedded in a chair \cite{chair_hrm}. Other potential applications of smart chair include automating the attendance process \cite{smart_chair_1} and self-position in a conference room \cite{chair_conference}.
In terms of eSports, there are few papers reporting on the cyber athlete performance estimation. In \cite{skill_capture_shooter} the authors perform the analysis of the dependence between the first- person shooter player skill and keyboard log data, mouse log data and game events. Other work related to the first-person shooter \cite{player_team_performance_study} deals with the personal player statistics based on the following metrics: KDA (Kills, Deaths, Assists) and Win/Loss rate.
Another set of works \cite{skill_dota_2_1, skill_moba_1, skill_moba_2} presents the research on how to reveal the most important factor of the player's skill in the MOBA (Multiplayer Online Battle Arena) genre. An analysis of the player performance in MMORPG (Massively Multiplayer Online Role-Playing Game) games is available in \cite{performance_motivation_mmorpg, modeling_mmorpg}. However, all of these papers focus on in-game and closely related data analysis without considering the physical condition of a player.

The contribution of this paper is the collection of eSports related physiological data and their further analysis using machine learning algorithms. As a result, we build the machine learning models and classify the data. The best model demonstrates 0.86 AUC in a binary classification. The novelty of this work is the experimentation with the professional eSports athletes and amateur players all specializing in CS:GO discipline.

This paper is organized as follows: in Section~\ref{Methodology} we present the methodology used in this work, in Section~\ref{Smart Chair Sensing Platform} we describe the smart chair platform used for data collection from the professional CS:GO athletes and amateur players. Machine learning algorithms are explained in Section~\ref{Machine Learning}. We discuss our experimental results and provide concluding remarks in Section~\ref{Evaluation} and Section~\ref{Conclusions}, respectively.

\section{Methodology}\label{Methodology}

In this research work 19 participants take part in the experiment: 9 professional athletes, primarily from the  Monolith professional team specializing in CS:GO discipline,  and 10 amateur players. Before the experiment, all the participants were informed about the project and the experiment details. On receiving written consents from the players, we conduced a questionnaire study to ensure that all the participants are in good form and do not take any drugs which could affect the experimental results.

In this work we consider the Retake modification of CS:GO discipline. In the \textit{Retake} scenario a terrorist team is opposed to a counter-terrorist team. The terrorist team is made up of 2 players who typically play in a defensive manner as they have a bomb planted on the territory and have to defend it from the opposite team. The counter-terrorist team including 3 players has the goal to deactivate the bomb or, alternatively, to kill the opposite team. The game user interface shows the bomb location on the map in the beginning of each round which lasts for approximately 40 s (there are 12 rounds in total). Players have to buy the same set of weapons for each round. The Retake scenario must be played continuously without any breaks between the rounds.

\section{Smart Chair Sensing Platform}\label{Smart Chair Sensing Platform}

\subsection{Smart Chair Design}\label{SmartChairDesign}

\begin{figure}[!b]
	\centerline{\includegraphics[width=\linewidth]{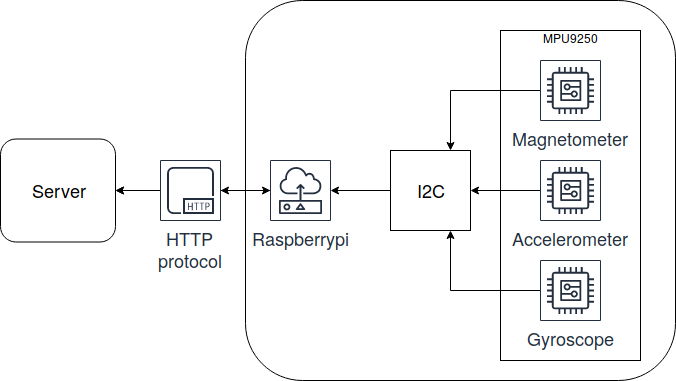}}
	\caption{Block diagram of proposed platform.}
	\label{scheme}
\end{figure}

While designing the smart chair we rely on the Internet of Things (IoT) paradigm~\cite{iot-2,iot1} - a number of sensors~\cite{sensors} are connected to a global network with intelligent capabilities. The smart chair platform consists of two units: a sensing unit  for data collection and a server for data processing. The block diagram of the platform is shown in Fig.\ref{scheme}.

\begin{figure}[!b]
	\centering
	\begin{subfigure}[]{0.18\textwidth}
		\centerline{\includegraphics[height=4.1cm]{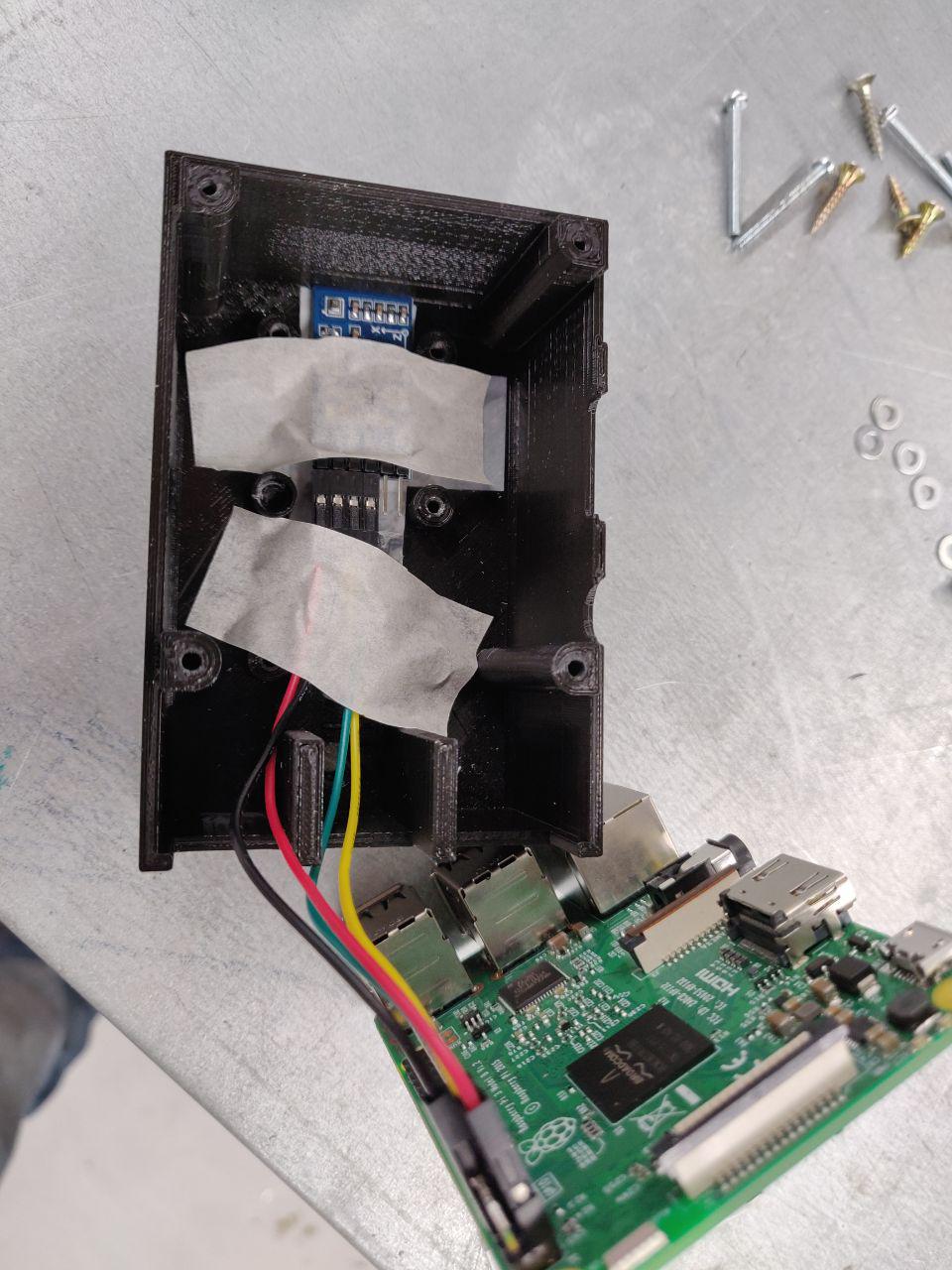}}
		\caption{Anatomy of sensing system.}
		\label{sensors}
%		\vspace{0.1cm}
	\end{subfigure}
	\begin{subfigure}[]{0.3\textwidth}
		\centerline{\includegraphics[height=4.1cm]{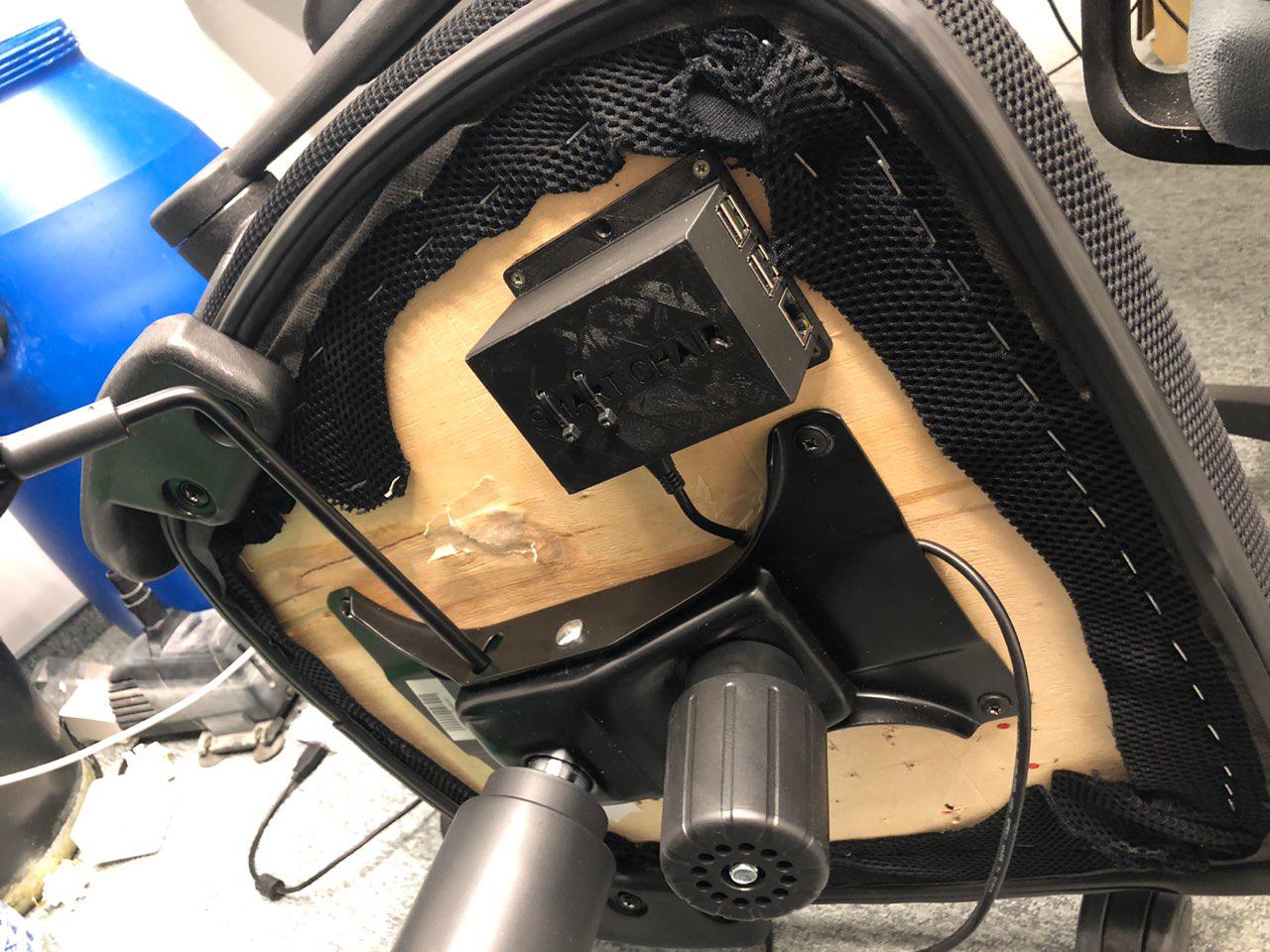}}
		\caption{sensing module fixed on the bottom of a chair.}
		\label{chair}
	\end{subfigure}
	\caption{Smart chair experimental testbed.}
	\label{sensing_system}
\end{figure}

The core element of the sensing unit is RaspberryPi-3 with Motion Processing Unit (MPU) 9250 which collects the data from the accelerometer, magnetometer and gyroscope via I2C protocol \cite{mpu_9250}.
The data are collected every 0.01 s. This solution can be further extended by adding extra sensors, e.g. pressure and temperature ones. Upon collecting the data, the sensing unit makes a request and sends it to a server over the WiFi wireless channel. The data are sent to the server via HTTP protocol every second in the JSON format. Data processing is realized on the server using the machine learning algorithms described in more details in the next section. All the machine learning algorithms are implemented using Python programming language. The smart chair experimental testbed is shown in Fig.~\ref{sensing_system}. It shoukd be noted here that the sensing unit is fixed at the bottom of a chair.

The cloud is an application, the front-end between the database, microcontroller and machine learning algorithms. The cloud application is developed in Java with Spring framework \cite{spring_framework}. This program allows us to run the application in different types of a cloud platform, for example, Cloud Foundry, Hysterix, Amazon Cloud \cite{spring_boot}. Also, Spring Data has an opportunity of using access technologies in different databases. The platform, therefore, can be connected to MySQL, PostgreSQL, Oracle databases, MongoDB \cite{spring_data}.

\subsection{Data Collection and Preprocessing}\label{DataCollection and Pre-processing}

The data collection procedure is crucial for further data analysis. It is essential to properly markdown the data for future convenient preprocessing.

In this experiment we collect the data from 19 persons playing CS:GO. Each CS:GO session lasts around 35 minutes and the sesnsors’ data are collectes every 0.01 s.

In our case we have 3 types of data. The accelerometer  provides acceleration, the magnetometer provides the angles between the sensor own axes X, Y, Z and the lines of force of the Earth magnetic field, the gyroscope provides the angular velocity. In this work, we present the data as parts of time series.

\begin{figure}[!bt]
	\centerline{\includegraphics[width=\linewidth]{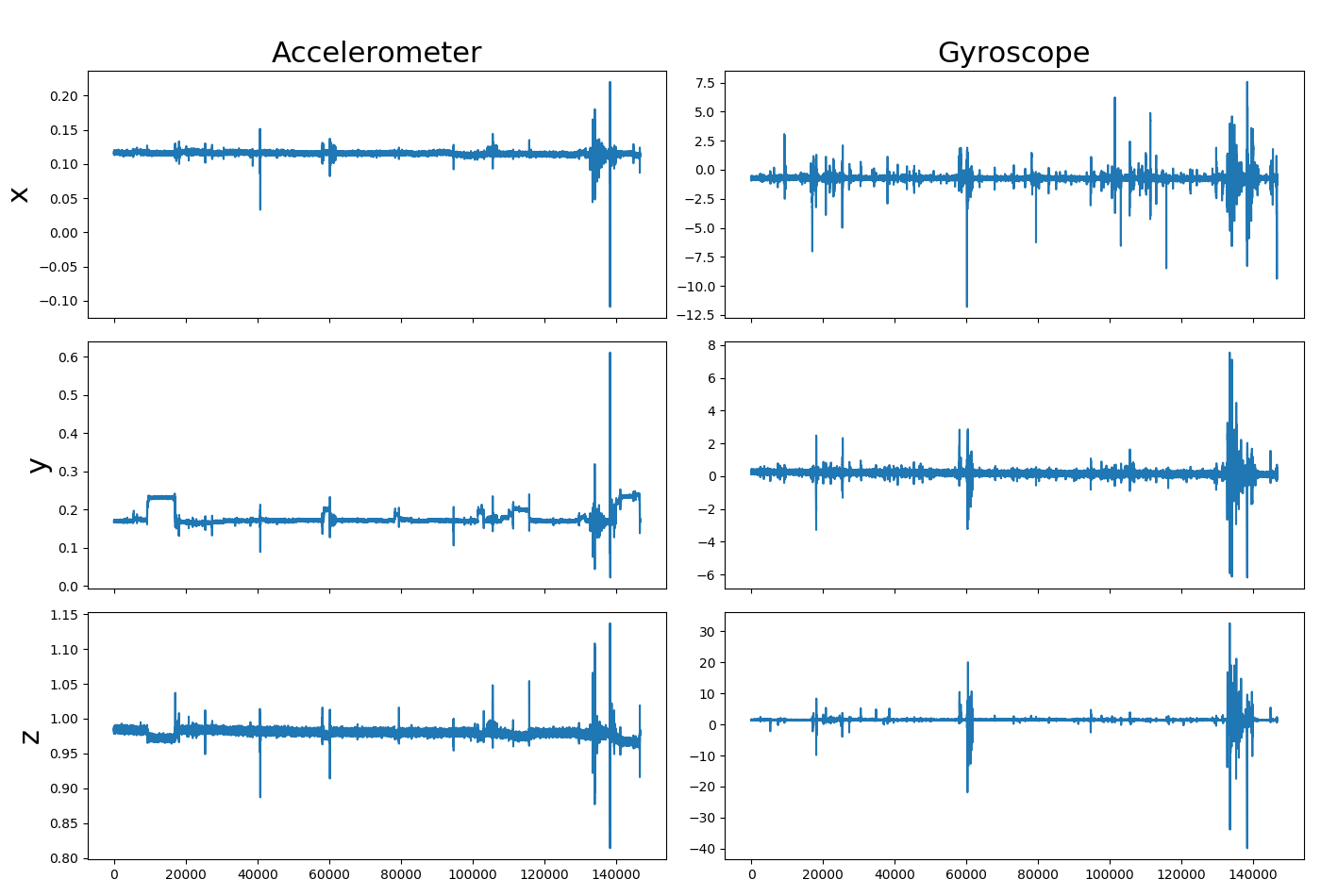}}
	\caption{Example of raw data collected from the accelerometer and gyroscope.}
	\label{raw_data}
\end{figure}

Example of raw data from the accelerometer and gyroscope are shown in Fig.~\ref{raw_data}. Axis \textit{z} corresponds to the vertical axis, \textit{y} corresponds to the axis passing through the player and the monitor, \textit{x} corresponds to the axis parallel to the gaming table.

We created two main feature groups based on the raw data. The first group is the portion of time when the data from thr sensor are deviating from the mean for more than 3 standard deviations. This means that a person is actively changing the position and moving. It can be named the active movement and considered for \textit{x}, \textit{y}, \textit{z} components of the accelerometer and gyroscope, resulting in 6 features in total.

The second group of six features is the mean dispersion of the sensors’ data when a person is not actively moving. It corresponds to the subtle oscillations by foot, head or other parts of the body. Besides, it is able to catch the slow and constant rotations on the chair.

An additional feature we used is the portion of time when the person leans on the back of the chair. It is extracted from the accelerometer data: if \textit{z}-coordinate is lower than a threshold, then the angle between \textit{z} and the vertical axes is too low and the person is leaning on the back of the chair.

\section{Machine Learning}\label{Machine Learning}

\subsection{Training Data}
We asked 19 participants (9 professional athletes and 10 amateur players) to estimate their skill against the low/high scale. After encoding the low skill to \textit{0} and the high skill to \textit{1} we got the binary target for machine learning models.

\begin{figure}[!bt]
	\centerline{\includegraphics[width=\linewidth]{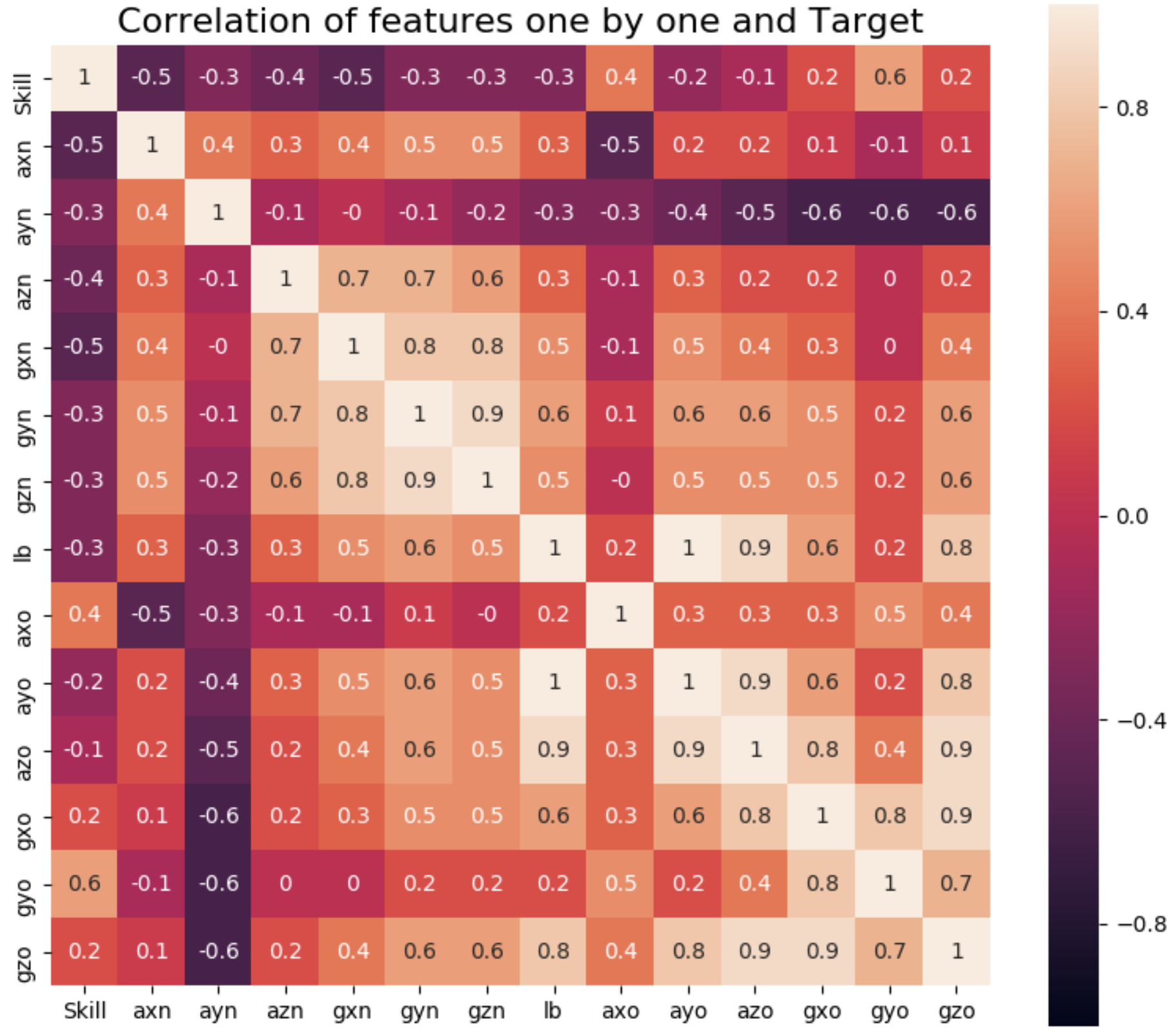}}
	\caption{Heat map for features.}
	\label{heatmap}
\end{figure}

From preprocessing step (see Section III-B) we have 13 features. Correlations between them and the target are represented in Fig.~\ref{heatmap}. The detailed description is provided in Table~\ref{heatmap_table}.

\begin{table}[]
	\caption{Description of features on the heat map.}
	\label{heatmap_table}
	\begin{center}
		\begin{tabular}{|p{0.6cm}|p{7.4cm}|}
			\hline
			\textbf{Name}  & \textbf{Description}                                                                        \\\hline
			Skill & Player’s skill in CS:GO. 0 or 1 (low or high)                                      \\\hline
			axn   & Active movement along x-axis. Intensity of rectilinear motion to the right/left.   \\\hline
			ayn   & Active movement along y-axis.Intensity of rectilinear motion to/from table.        \\\hline
			azn   & Active movement along z-axis. Intensity of rectilinear motion up/down.             \\\hline
			gxn   & Active rotation on x-axis. Frequency of approaching to/distancing from a monitor.  \\\hline
			gyn   & Active rotation on y-axis. Intensity of swaying to the right/left side of a chair. \\\hline
			gzn   & Active rotation on z-axis. Intensity of rotations on a vertical axis.              \\\hline
			lb    & Portion of time when player leans to the back of a chair.                          \\\hline
			axo   & Intensity of subtle rectilinear oscillations parallel to table.                    \\\hline
			ayo   & Intensity of subtle rectilinear oscillations to/from table.                        \\\hline
			azo   & Intensity of subtle rectilinear oscillations up/down.                              \\\hline
			gxo   & Intensity of subtle approaching to/distancing from a monitor.                      \\\hline
			gyo   & Intensity of subtle swaying to the right/left side of a chair.                     \\\hline
			gzo   & Intensity of subtle rotations on a vertical axis.                                 \\\hline
		\end{tabular}
	\end{center}
\end{table}

We observe that the professional athletes perform less active movements during the game (rows/columns 2-7). Less intensive active movements could be connected with the higher concentration on the game. On the other hand, some subtle motions (rows/columns 9-14) are more characteristic of the athletes. It includes the rotational motion on the y-axis and the rectilinear motion along the x-axis. The first corresponds to the point on how much the person is swaying from right to left; the second means that the person moves in rectilinear way from right to left. Leaning on the back of the chair is more representative of the amateur players.

In order to increase the amount of data we divided the log of  each player into 3-minutes sessions. As a result, our dataset includes 154 sessions from 19 persons. We have fitted several machine learning models based on 13 features to predict the player skill.

\subsection{Machine Learning Algorithms}\label{ml_algorithms}
\subsubsection{Logistic regression}

It is the classifier which uses the logistic function to calculate probability from margin and maximizing the likelihood function,  
$L$, % w.r.t. parameters $\theta$
given the observations \cite{ng_lectures}:

\begin{equation}
L(\theta) = \text{Pr} (Y|X; \theta) = \prod_i \text{Pr}(y_i | x_i; \theta),
\label{lr}
\end{equation}
where $X$ is a design matrix, $Y$ is a vector of targets, $\theta$ is the vector of model parameters. Probability of being from class 1 is determined by sigmoid function \eqref{sigmoid}:

\begin{equation}
\text{Pr}(y_i | x_i; \theta) = \frac1{1 + e^{-\theta^\top x_i}}.
\label{sigmoid}
\end{equation}

In our experiment the main advantages of the logistic regression are stability, easy interpretation and good approximation.

\subsubsection{Support Vector Machine (SVM)}

We also used SVM classifier with soft-margins \cite{svm_classic}. This method tries to separete classes by a hyperplane so that the gap between them is maximum possible:

\begin{align}
\label{eqn:eqlabel}
\begin{split}
\min_{w, \xi}~& \frac12 w^\top w + \frac\gamma{2}\sum_ {i=1}^n \xi_i,
\\
\text{subject to~}&y_i(w^\top x_i + b) \geq 1 - \xi_i,\\
& \xi_i \geq 0,  i \in 1 \dots n,
\end{split}
\end{align}
where $x_i$ is a feature vector, $y_i$ is a scalar target, $w$ and $b$ are the parameters determining the separation hyperplane and its width, $\xi_k$ is a slack variable, $\gamma$ is the parameter which determines the tradeoff between the maximum margin and the minimum classification error.

The main advantage of this method is, again, stability due to  maximization of the gap between different classes.

\subsubsection{Nearest neighbors}

For the given input, $k$-nearest neighbors classifier searches for the $k$ nearest neighbors in the feature space from the training set and returns the most popular label among them \cite{cover1967nearest}. In our problem we used the number of neighbours equal to 5, which provides the maximum ROC AUC.

In our case this simple algorithm can predict the unknown player performance taking into account performance of similar players.

\subsubsection{Random Forest}

A random forest is the classifier represented by an ensemble of tree-structured classifiers. Random forest can handle complex dependencies in data, but this may lead to overfitting. According to experiments the optimal maximum tree depth in our problem is 4. 

For our problem random forest can learn logical rules to distinguish the high-skilled and low-skilled players and catch more complex patterns in their behaviours.

\section{Evaluation}\label{Evaluation}

We can not use sessions from the same player for both training and testing stage, because our model should be able to predict the performance for new participants.
Thus, to estimate the model performance correctly we trained models, described in \ref{ml_algorithms}, on all people except 4-5 out of 19 and then validated on them. For more stable results the scores were calculated 100 times and averaged.

We used ROC AUC score \cite{fan2006understanding} as an evaluation metric, for it nicely represents how well the  classes are separated by a model. The maximum possible value is 1, while the random guess get 0.5. Scores for different algorithms are shown in Table~\ref{table_scores}. Linear models, such as Logistic Regression and SVM perform better than KNN and Random Forest. It can be explained that the real dependence between the player skill and extracted features is similar to the linear dependency.
Besides, it is possible that KNN and Random Forest are too complex for our problem, for they tend to overfit for small data.

\begin{figure}[!tb]
	\centerline{\includegraphics[width=\linewidth]{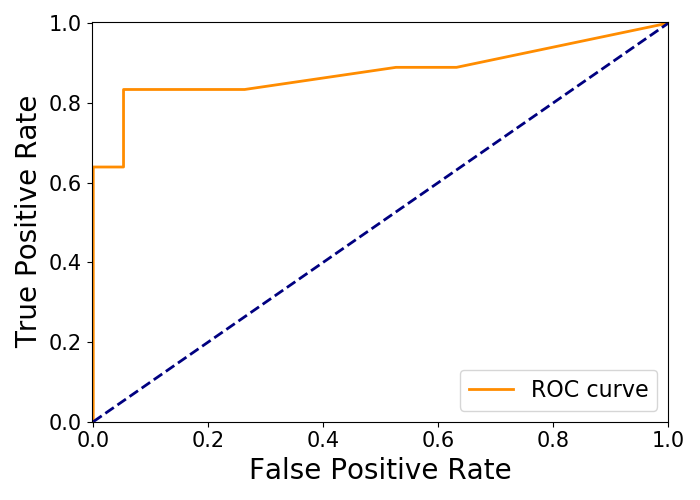}}
	\caption{ROC AUC curve for logistic regression.}
	\label{roc_auc}
\end{figure}

The mean ROC AUC score for all of the algorithms is more or equal to 0.8, which means that the eSport athlete performance can be successfully predicted by machine learning models. As an illustrative example, ROC curve for Logistic Regression is shown in Fig.~\ref{roc_auc}.

\begin{table}[!bt]
	\caption{Models performance on predicting the player skill: a comparative study.}
	\label{table_scores}
	\begin{center}
		\begin{tabular}{|l|l|l|}
		\hline
		\textbf{Method}                 & \textbf{AUC, mean} & \textbf{AUC, std} \\\hline
		Logistic Regression    & 0.85      & 0.14     \\\hline
		Support Vector Machine & \textbf{0.86}      & 0.13     \\\hline
		KNN, 5 neighbours      & 0.80      & 0.13     \\\hline
		Random Forest, depth 4 & 0.82      & 0.16  \\\hline  
		\end{tabular}
	\end{center}
\end{table}

\begin{figure*}[!hbt]
	\centerline{\includegraphics[width=\linewidth]{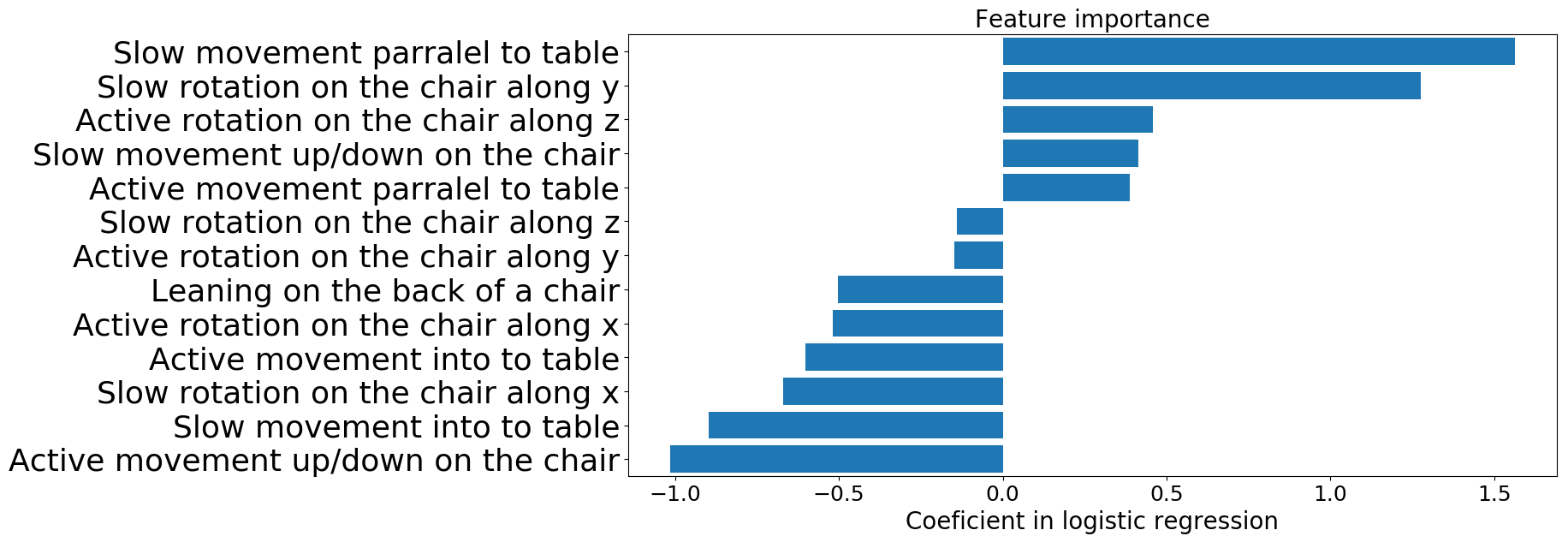}}
	\caption{Feature importance as coefficients in the logistic regression.}
	\label{feature_importance}
\end{figure*}

To figure out which characteristics define the player performance we used coefficients in the logistic regression. It is a stable and reliable estimation of feature importance. Positive coefficients correspond to the high-skilled players behavior, while negative coefficients are typical for the  low-skilled players behavior. Large absolute value means that feature is more important. Fig.~\ref{feature_importance} demonstrates that the motion to right and to left is the most characteristic of professional athletes. Active rotations on the chair are also important factors, but typically refer to the amateur players.
These results are consistent with the recap from the heatmap shown in Fig.~\ref{heatmap}.

%\begin{figure*}[!bt]
%	\centerline{\includegraphics[width=\linewidth]{pic/heatmap_last.png}}
%	\caption{Correlations between features.}
%	\label{heatmap_features}
%\end{figure*}

\section{Conclusions}\label{Conclusions}

In this work, we have proposed the IoT based platform for data collection and analysis in the context of eSports. The system consists of a sensing unit with the sensors integrated in a chair, and a server which perform the collected data analysis based on machine learning algorithms. In particular, we assess the professional level of the players in CS:GO discipline.

We have demonstrated that Machine Learning model is able to predict the professional level of the person playing a quick Retake match in CS:GO. This approach brought ROC AUC 0.86 score for test data from the support vector machine for the binary classification. Most important factors have been figured out.

In our future work we are intending to increase the number of and add some other types of sensors on the platform. Thus, the more diverse features can be generated and more complicated models can be applied. We are also planning to invite more participants to the experiment.

For securing more accurate feature extraction from time-series we suggest to use the automatic segmentation by hidden Markov models \cite{Market2009}, non-parametric online anomaly detection to pre-process data \cite{ConformalAD2015} and ensembles of detectors for online prediction of eSports athletes performance \cite{EnsemblesDetectors2015,AggregationLongTerm}.

\section*{Acknowledgment}

The reported study was funded by RFBR according to the research project No. 18-29-22077\textbackslash18.

Authors would like to thank Skoltech Cyberacademy, CS:GO Monolith team and their coach Rustam “TsaGa” Tsagolov for fruitful discussions while preparing this article. Also, the authors thank Alexey "ub1que" Polivanov for supporting the experiments by providing a slot at the CS:GO Online Retake server (http://ub1que.ru).

\bibliography{main}{}
\bibliographystyle{IEEEtran}

\end{document}